\documentclass[11 pt]{article}
\usepackage{amsmath, mathtools}
\usepackage{amsthm}
\usepackage{amssymb}
\usepackage{graphicx}
\usepackage{appendix}
\usepackage{color}
\usepackage{hyperref}
\usepackage{wrapfig}

\usepackage{float}

\usepackage{multirow}

\numberwithin{equation}{section}

\begin{document}

\title{A data-science-driven short-term analysis of Amazon, Apple, Google, and Microsoft stocks}
\author{Shubham Ekapure\footnote{ Author ordering is based on the alphabetical order of the last names. Contributions of the first three authors are equal.  The last author is the research mentor of this REU project. }    \footnote{Indian Institute of Technology Kharagpur, West Bengal, India}, Nuruddin Jiruwala\footnote{Indian Institute of Technology Kharagpur, West Bengal, India}, Sohan Patnaik\footnote{Indian Institute of Technology Kharagpur, West Bengal, India},  Indranil SenGupta\footnote{Corresponding author. Associate Professor and Graduate Program Director, Department of Mathematics, North Dakota State University, Fargo, North Dakota, USA. Email: indranil.sengupta@ndsu.edu}}
\date{\today}
\maketitle

\begin{abstract}

In this paper, we implement a combination of technical analysis and machine/deep learning-based analysis to build a trend classification model. The goal of the paper is to apprehend short-term market movement, and incorporate it to improve the underlying stochastic model. Also, the analysis presented in this paper can be implemented in a \emph{model-independent} fashion. We execute a data-science-driven technique that makes short-term forecasts dependent on the price trends of current stock market data. Based on the analysis, three different labels are generated for a data set: $+1$ (buy signal), $0$ (hold signal), or $-1$ (sell signal). We propose a detailed analysis of four major stocks- Amazon, Apple, Google, and Microsoft. We implement various technical indicators to label the data set according to the trend and train various models for trend estimation. Statistical analysis of the outputs and classification results are obtained.
   
\end{abstract}

\textsc{Key Words:}  Short-term forecasting, stochastic model, feature extraction, classification results, LSTM.


\section{Introduction}

Stock market prediction is the act of trying to determine the future value of a company stock or other financial instrument traded on an exchange. The successful prediction of future price of a stock could yield significant profit. A long-term model for a stock is ambitious and may not be very realistic. However, there have been attempts to model the stock for a short-term. There has been a fair amount of research in this field and the prediction methodologies can be broadly classified into three categories- fundamental analysis, technical analysis, and machine/deep learning-based analysis.

The fundamental analysis in the stock market aims to estimate the true values of the stock price. After that, the stock price can be compared with its traded value on the stock markets. Thus, this analysis finds out whether the stock on the market is undervalued or not (see \cite{fund}). Finding out the true value can be performed by various methods with basically the same principle. The principle is that a company is worth all of its future profits added together. These future profits also must be discounted to their present value. This principle goes along well with the theory that a business is all about profits and nothing else. As a matter of fact, fundamental analysis is a long-term strategy that indeed, makes the stock trading decision difficult in the short run. In the literature, the fundamental analysis is also implemented to examine various financial statements with the aim to assess a real value of company's stock. The paper \cite{fund2} provides a systematization of the fundamental analysis.

Researchers, especially the short-term traders, are interested in the short-term forecasting of the stock price and estimating the trend. Due to the dynamic nature and volatility in stock price, even short-term prediction of the stock price becomes a challenging task. As technology is continually improving, stock traders tend to move towards using \emph{intelligent trading systems} rather than fundamental analysis for predicting prices of stock. This helps the traders make immediate investment decisions. This leads to the next two categories, technical analysis and machine/deep learning-based analysis, mentioned above.

The technical analysis seeks to determine the future price of stock based solely on the trends of the past price (see \cite{tech}). Some of the common technical indicators that are used to estimate the stock price movements are: exponential moving average (EMA), Bollinger bands, moving average convergence divergence (MACD), momentum, volatility, RSI index etc. Technical analysis is used for short-term strategy plans, therefore, making it relevant to the field of short-term market prediction.

Finally, in the category of machine/deep learning-based analysis, a good deal of methods are implemented in the recent literature. For example, in \cite{jiang}, certain index options are constructed by efficient algorithms including uniform approximation error.  In \cite{mome2} the authors study the optimal timing for an asset sale for an agent with a long position in a momentum trade. In \cite{mome} a ``deep momentum network" is introduced  which is a hybrid approach that incorporates deep learning-based trading rules into the volatility scaling framework of time series momentum. The model also simultaneously learns both trend estimation and position sizing from the empirical data. In \cite{AODS} the authors propose a refinement of stock price dynamics over some existing stochastic model. This refinement is obtained through the application of data-science, especially machine/deep learning algorithms. This machine/deep learning-based redefined model is implemented for commodity markets in \cite{Roberts1, Roberts2}.

In this paper, we implement a combination of technical analysis and machine/deep learning-based analysis in order to build a trend classification model. That means, for the model and analysis proposed in this paper, in addition to the robust artificial intelligence (AI) model, we use technical indicators that capture the stock trend movement with greater accuracy. The rest of the paper proceeds as follows. In Section \ref{sec2}, we provide a possible improvement of the underlying stochastic model based on the data-science-driven short-term forecasting of the stock prices. In Section \ref{sec3}, we provide a detailed description and perform exploratory data analysis of the data sets that are considered for this paper. The methodology for a data-science-driven short-term forecasting is provided in Section \ref{sec4}. Statistical analysis of the outputs and classification results are provided in Section \ref{sec5}. Finally, a brief conclusion is provided in Section \ref{sec6}. Some relevant statistical quantities for the data sets are provided in Appendix \ref{apendixA}.

\section{Mathematical formulation}
\label{sec2}

We assume that the share price $S_t$ is given by
\begin{equation}
\label{main1}
S_t= S_0e^{X_t}, \quad \text{where} \quad
dX_t= b_t\,dt +  \sigma_t dW_t+ \theta_t  dJ_t,
\end{equation}
where $b_t$ is a deterministic function of $t$, $W_t$ is the Brownian motions, and $J_t$ is the jump process with intensity $\lambda$.  We assume that $W_t$ and
$J_t$, are independent. 
We write $J_t$ in terms of integral with respect to Poisson random measures $N(dt, dx)$. Consequently,
\begin{equation*}
  J_t= \int_0^t \int_{\mathbb{R}} x N(dt,dx).
\end{equation*}
Hence  \eqref{main1} can be written as 
\begin{equation}
\label{main2}
S_t= S_0e^{X_t}, \quad \text{where} \quad
dX_t= b_t\,dt +  \sigma_t dB_t+  \theta_t \int_{\mathbb{R}} x N(dt,dx).
\end{equation}
In addition to that, $\sigma_t$  is assumed to be stochastic, and its dynamics are governed by
\begin{equation}
\label{vol}
d\sigma_t^2= F(\sigma_t^2, \beta_t H_t)
\end{equation}
for an appropriate function $F$, where $H_t$ is a jump process with intensity $\mu$.

There are some special cases of the proposed model that are studied  in literature in connection with the financial market, such as the Barndorff-Nielsen and Shephard (BN-S) model. For the BN-S model, the share price (see , \cite{BN-S1, BN-S2, NV}) or commodity price (see, \cite{sub1, sub2}) $S_t$  on some risk-neutral probability space $(\Omega, \mathcal{G}, (\mathcal{G}_t)_{0 \leq t \leq T}, \mathbb{Q})$ is modeled by

\begin{equation}
\label{1}
S_t= S_0 \exp (X_t),
\end{equation}
\begin{equation}
\label{2}
dX_t = (B- \frac{1}{2} \sigma_t ^2 )\,dt + \sigma_t\, dW_t + \rho \,dZ_{\lambda t}, 
\end{equation}
\begin{equation}
\label{3}
d\sigma_t ^2 = -\lambda \sigma_t^2 \,dt + dZ_{\lambda t}, \quad \sigma_0^2 >0,
\end{equation}
where the parameters $B \in \mathbb{R}$, $\lambda >0$ and $\rho \leq 0$.  In the above model, $W_t$ is a Brownian motion, and the process $Z_{\lambda t}$ is a subordinator. Also $W$ and $Z$ are assumed to be independent, and $(\mathcal{G}_t)$ is assumed to be the usual augmentation of the filtration generated by the pair $(W, Z)$. We consider a special case of \eqref{main2}, where $dZ_{s}= \frac{1}{\rho} \int_0^{\infty} x N(ds, dx)$, is a subordinator.  
Making a scaling in the time variable, we define $s= \lambda t$, for $\lambda>0$. Then, we obtain,
$dZ_{\lambda t}= \frac{1}{\rho} \int_0^{\infty} x N(\lambda \,dt, dx)$. 

Since $ \rho \leq 0$, from \eqref{1} and \eqref{2}, it is clear that the value of $S_t$  primarily grows with the drift coefficient $B$, and decays with the action of the subordinator $Z$. 
For a short-term forecasting, we re-frame \eqref{2} as follows:
\begin{equation}
\label{2111}
dX_t = (B(\theta)- \frac{1}{2} \sigma_t ^2 )\,dt + \sigma_t\, dW_t + \rho(\theta) \,dZ_{\lambda t}, 
\end{equation}
where $\theta$ can take three values, $0$ and $\pm1$, and for an appropriate $\Lambda \in \mathbb{R}$ (obtained from data calibration),

\begin{equation}
\label{Btheta}
  B(\theta) =
  \begin{cases}
                                   B, & \text{if $\theta= 0, -1$}, \\
                                   B+ \Lambda^2, & \text{if $\theta =1$}, \\
 \end{cases}
\end{equation}
and
\begin{equation}
\label{rhotheta}
  \rho(\theta) =
  \begin{cases}
                                   \rho, & \text{if $\theta= -1$}, \\
                                   0, & \text{if $\theta =0, 1$}. \\
 \end{cases}
\end{equation}

Consequently, the indicator function $\theta$, that provides a \emph{signal} for the short-term share price movement, can generate three scenarios: \\
\textbf{Case 1}: $\theta=1$, where a short-term upward share movement is expected:
\begin{equation*}
dX_t = (B+ \Lambda^2- \frac{1}{2} \sigma_t ^2 )\,dt + \sigma_t\, dW_t . 
\end{equation*}
\textbf{Case 2}: $\theta=0$, where not much short-term share movement is expected:
\begin{equation*}
dX_t = (B- \frac{1}{2} \sigma_t ^2 )\,dt + \sigma_t\, dW_t . 
\end{equation*}
\textbf{Case 3}: $\theta=-1$, where a short-term downward share movement is expected:
\begin{equation*}
dX_t = (B -\frac{1}{2} \sigma_t ^2 )\,dt + \sigma_t\, dW_t +  \rho \,dZ_{\lambda t}. 
\end{equation*}
In summary, the share price model, given by \eqref{1}, \eqref{2111}, and \eqref{3}, is driven by an indicator function $\theta$ that can take values $1$, $0$, or $-1$, for short-term anticipated upward, stable  or downward share movements, respectively. For the rest of the paper, we derive a data-science driven approach for finding the appropriate indicator function. In effect, the new model re-frames and improves the classical BN-S model.

\section{Data sets}

\label{sec3}

We consider four stock data sets, viz. Amazon, Apple, Google, and Microsoft, to carry out the statistical analysis. Training data sets are obtained between the years 2014 and 2019.  For testing, we consider intervals in the year 2021. Before executing the technical analysis on the data, we perform exploratory data analysis on each of the data sets. Three approaches as described below are primarily implemented for the basic data analysis. The primary objective is to get an estimation of the near-term stock fluctuation. For a \emph{model-independent analysis}, this is sufficient for a short-term investor to decide whether to sell, hold or buy a stock. From the perspective of a stochastic model (i.e., for a \emph{model-dependent analysis}),  such estimation will lead to the appropriate value of $\theta$ in \eqref{Btheta} and \eqref{rhotheta}.

\begin{itemize}
\item \textbf{t-SNE Clustering Analysis}: t-SNE clustering is a statistical method for visualizing high-dimensional data by giving each data point a location in a two or three-dimensional map. The idea behind t-SNE can be explained in simple terms: we are given with a set ${x_1, . . . , x_n} \in \mathbb{R}$, where $x_1, x_2, \dots$, are the feature vectors at a particular time step. These multi-dimensional features are used for t-SNE clustering which maps them down to 2 dimensions using the \emph{principal component analysis} (PCA). It is observed that the Google stock data is clustered into two groups with some data points (outliers) outnumbered by the other two clusters, whereas for Amazon, Apple, and Miscrosoft data, 3 prominent clusters are found. The number of clusters can be thought of furnishing an intuition regarding types of trends in the data. The time-series plot of Google data is somewhat stable, and it keeps on increasing. This leads to the idea of two class labels- ``increasing" and ``constant" (or, ``stable"). For the other three data sets, three class labels can be assigned corresponding to ``increasing", ``constant" (or, ``stable"), and ``decreasing" stock trends. Images of the Apple, Amazon, and Microsoft data sets, which are split into three clusters are shown in Figure 1. In addition, the image of the Google data set which is split into two clusters is shown in Figure 1.

\begin{figure}[H]
    \centering
        \begin{tabular}{| c | c |}
             \hline
              \includegraphics[scale = 0.45]{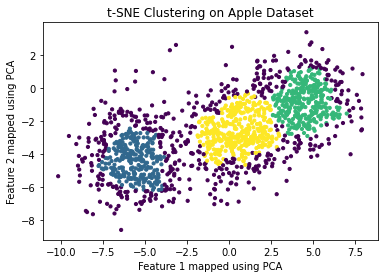} & \includegraphics[scale = 0.45]{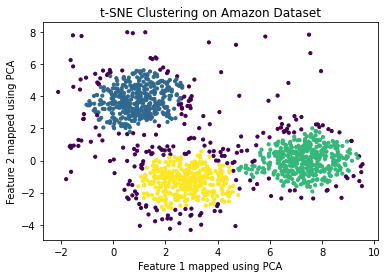} \\
             \hline
             \includegraphics[scale = 0.45]{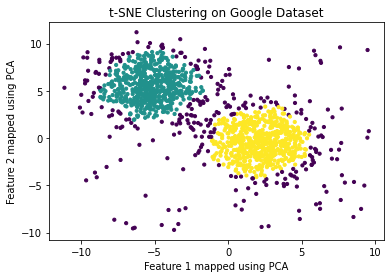} &  \includegraphics[scale = 0.45]{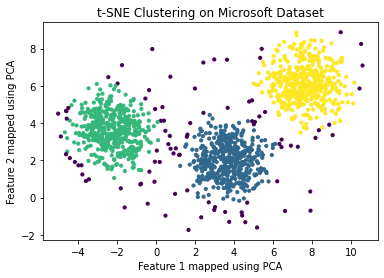} \\
             \hline
        \end{tabular}
\caption{t-SNE clustering for Apple, Amazon, Google, and Microsoft stocks.}
\end{figure}

\item \textbf{Autocorrelation and correlation}: Autocorrelation is a mathematical representation of the degree of similarity between a given time series and a lagged version of itself over successive time intervals. In all the stock data analysis for this paper, it is observed that the stock prices have a high correlation with the near future prices. Intuitively, this gives the idea that time series analysis models can be employed to estimate the stock price as well as capture the trends. The auto-correlation plot for the Apple data is provided in Figure 2. This shows that a high correlation is present when the lag is low (implying strong dependence on near future stock prices).
\begin{figure}[H]
\centering
\includegraphics[scale = 0.65]{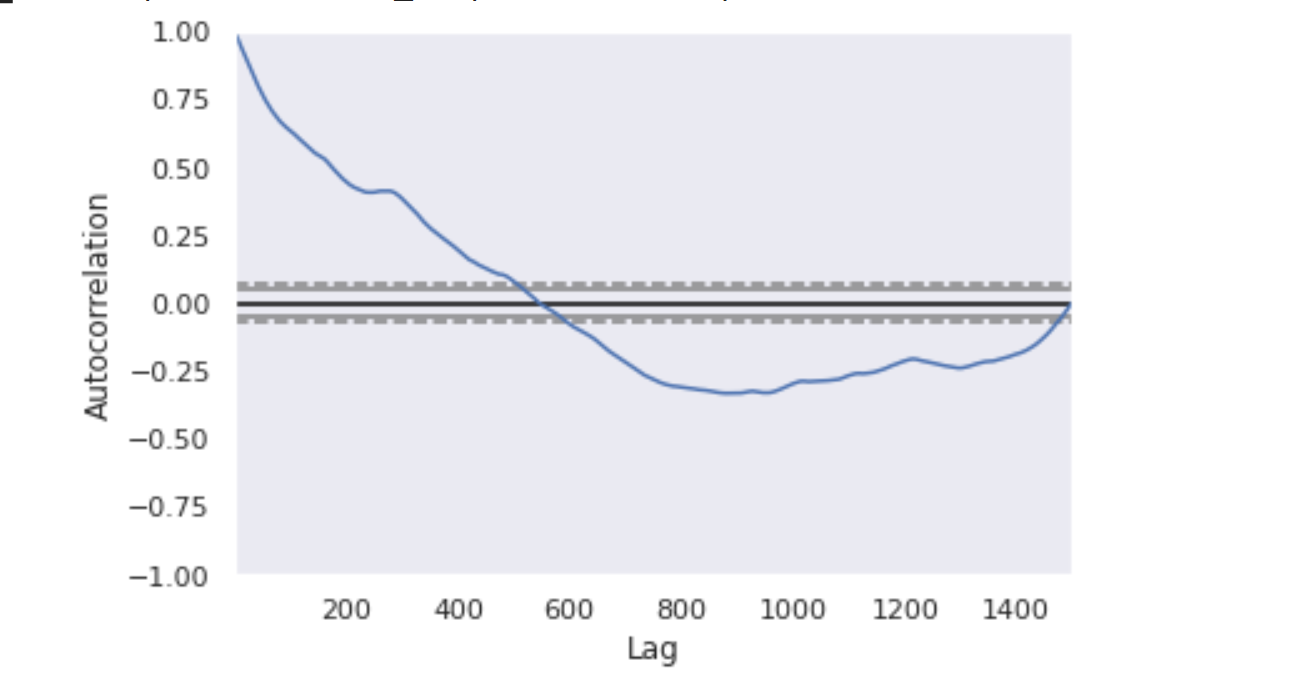}
\caption{Auto-Correlation plot for the Apple stock.}
\end{figure}
The autocorrelation plots for Amazon, Google and Microsoft stocks are shown in Figure 3, Figure 4, and Figure 5, respectively.
\begin{figure}[H]
\centering
\includegraphics[scale = 0.65]{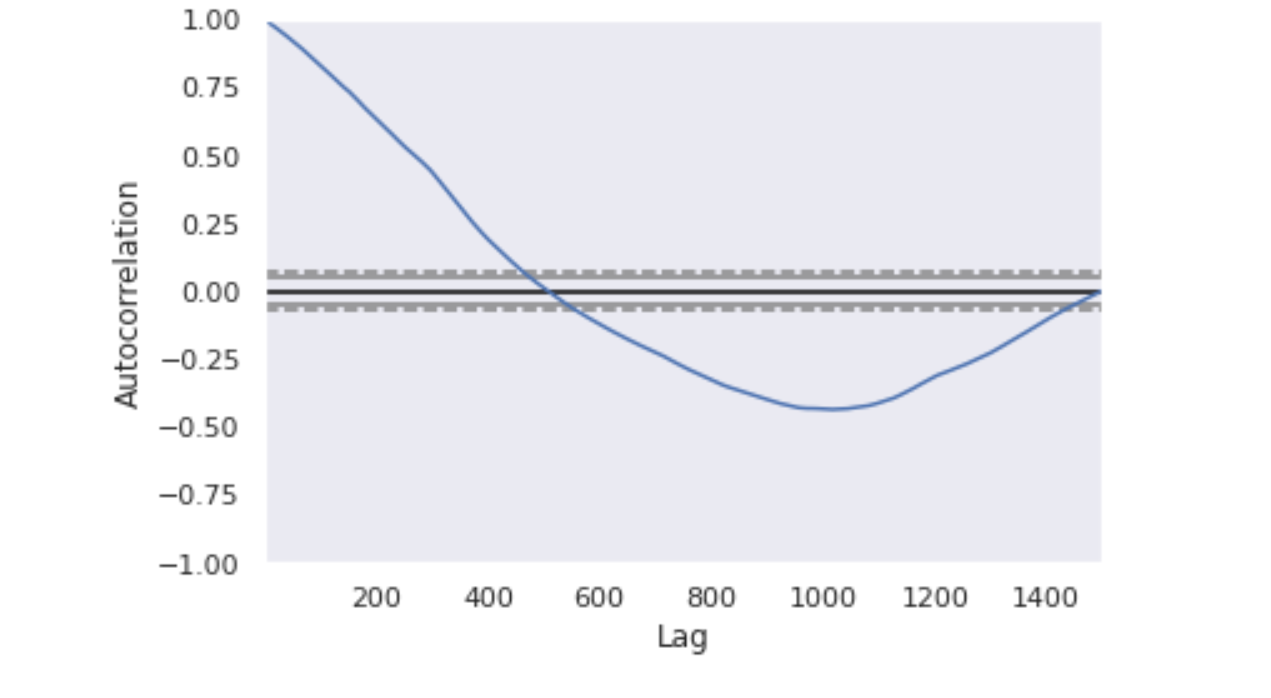}
\caption{Auto-Correlation plot for the Amazon stock.}
\end{figure}
\begin{figure}[H]
\centering
\includegraphics[scale = 0.65]{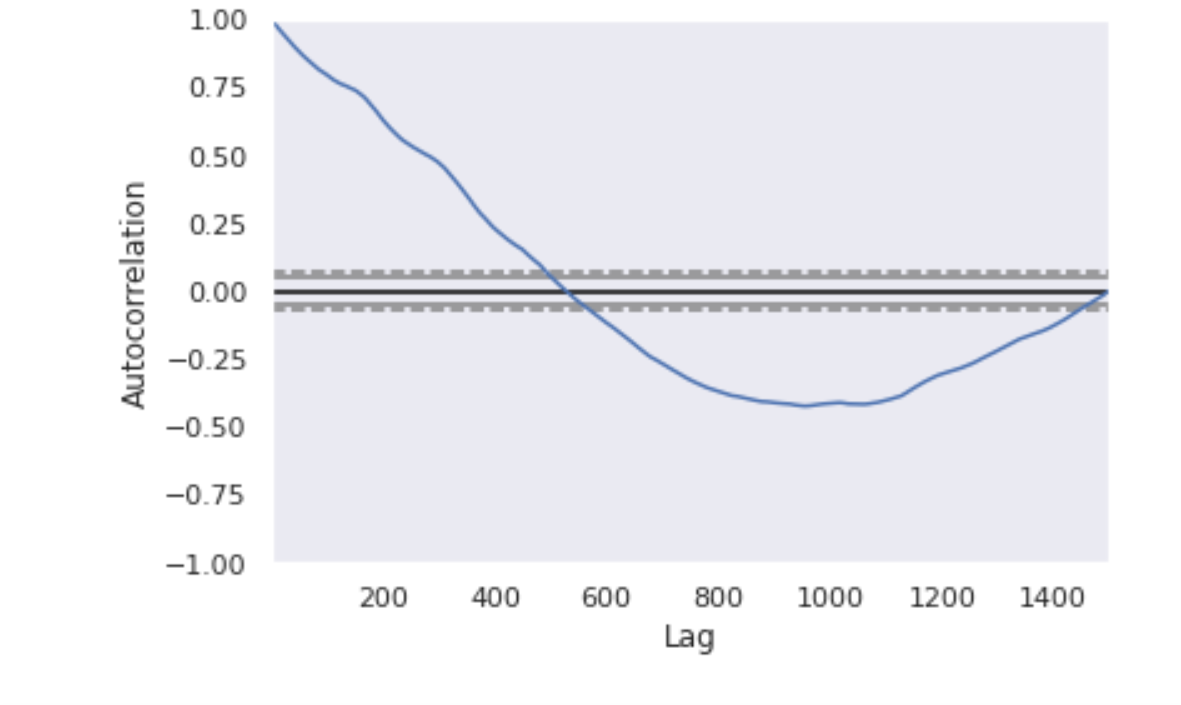}
\caption{Auto-Correlation plot for the Google stock.}
\end{figure}
\begin{figure}[H]
\centering
    \includegraphics[scale = 0.65]{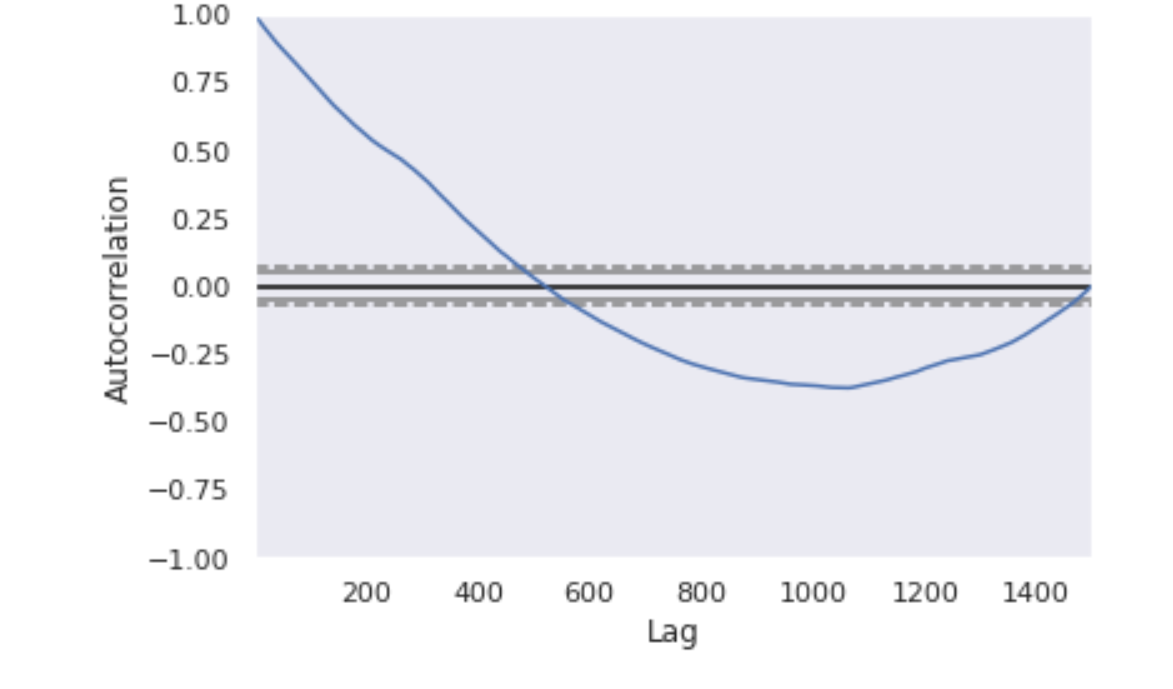}
\caption{Auto-Correlation plot for the Microsoft stock.}
\end{figure}

\item \textbf{PCA and Cosine Similarity}: The multi-dimensional features are labeled $+1$ or $-1$ using the indicators that are mentioned in the subsequent methodology section. Once the labeling of the data set is completed, the \emph{principal component analysis} (PCA) is implemented on the feature vectors followed by the computation of cosine similarity between all the pairs of feature vectors in each of the classes of labels $+1$ and $-1$. It is used to assess how different historical prices distribution vector are, and how related the feature vectors are for their respective classes (i.e., $+1$ and $-1$). We also note that the labels $+1$ and $-1$ in this context are the class labels of the stock trend- they do not stand for the similar or dissimilar classes.  \end{itemize}

Various statistical quantities are computed for each of the four data sets on the attributes, such as \emph{open price}, \emph{high price}, \emph{low price}, \emph{close price}, \emph{adjusted close price}, \emph{volume SMA close}, \emph{EMA close}, \emph{up}, \emph{down}, and \emph{RSI close}. The results are provided in Appendix \ref{apendixA}.

\section{Methodology}
\label{sec4}

\subsection{Feature extraction and  label generation}
\label{sec41}
 The technical indicators can be broadly classified into four categories, viz. based on trend indicator, momentum indicator, volume indicator, and volatility indicator. False signals arise when a trader uses two or more indicators of the same category in order to make a trading decision. For the technical analysis of this paper, we implement MACD which is based on the trend indicator; RSI and TRIX  which are based on the momentum indicator; and Bollinger Bands which is based on the volatility indicator. Hence, our choice of indicators for the classification pipelines successfully eliminates the false signal threat. The technical indicators are briefly summarized below:
    
\begin{itemize}
\item \textbf{MACD}:  The moving average convergence divergence (MACD) is a trend-following momentum indicator that shows the relationship between two moving averages of a security’s price (see \cite{macd}). The MACD is calculated by subtracting the 26-period exponential moving average (EMA) from the 12-period EMA. This causes MACD to oscillate around the zero level. A signal line is created with a 9 period EMA of the MACD line. The formula for calculating MACD is as follows:         
        \begin{equation}
           \text{MACD} = \text{EMA}_{12}(C) - \text{EMA}_{26}(C),
        \end{equation}
and
        \begin{equation}
           \text{Signal Line} = \text{EMA}_9(\text{MACD}),
        \end{equation}
where $C$ is the closing price, and $\text{EMA}_n$ is the $ n$-day exponential moving average. We consider the first instance of fall or rise of $\text{MACD}$ line relative to the $\text{Signal Line}$, and buy or sell accordingly. In other cases, we hold.

 \item \textbf{RSI}: The relative strength index (RSI) is a popular momentum indicator which determines whether the stock is overbought or oversold (see \cite{tech2}). A stock is said to be overbought when the demand unjustifiably pushes the price upwards. In effect, this implies that the stock is overvalued and should be sold. A stock is said to be oversold when the price goes down sharply to a level below its true value. RSI ranges from $0$ to $100$ and generally, when $\text{RSI} \geq 70$, it may indicate that the stock is overbought and when $\text{RSI} \leq 30$, it may indicate the stock is oversold. The formula for RSI is as follows:
          \begin{equation}
            \text{RSI} = 100 - \frac{1}{1 + \text{RS}},
        \end{equation}
where
        \begin{equation}
            \text{RS} = \frac{\text{Average Gain Over past 14 days}}{\text{Average Loss Over past 14 days}}.
        \end{equation}

\item \textbf{TRIX}: The triple exponential moving average oscillator (TRIX) is a momentum indicator that oscillates around zero. It displays the percentage rate of change between two triple smoothed exponential moving averages. The formula for TRIX is as follows:
        \begin{equation}
            \text{TRIX}(i) = \frac{\text{EMA}_3(i) - \text{EMA}_3(i-1)}{\text{EMA}_3(i-1)}.
        \end{equation}

\item \textbf{Bollinger Bands}:  The Bollinger bands (BBANDS) study plots upper and lower envelope bands around the price of the instrument (see \cite{bband}). The width of the bands is based on the standard deviation of the closing prices from a moving average of price.
\end{itemize}

Figure 6 is the visualization of the plots of technical indicators such as 7 and 21 day moving averages, closing price, upper and lower bands corresponding to Bollinger bands, momentum, MACD and logarithmic mapping of momentum. The logarithmic mapping provides a fair estimate of the stability of the stock time series data. Moreover, the Bollinger Bands provide an intuition about the deviation of the stock from the simple moving average of the closing price. 
    
\begin{figure}[H]
\centering
 \includegraphics[scale = 0.4]{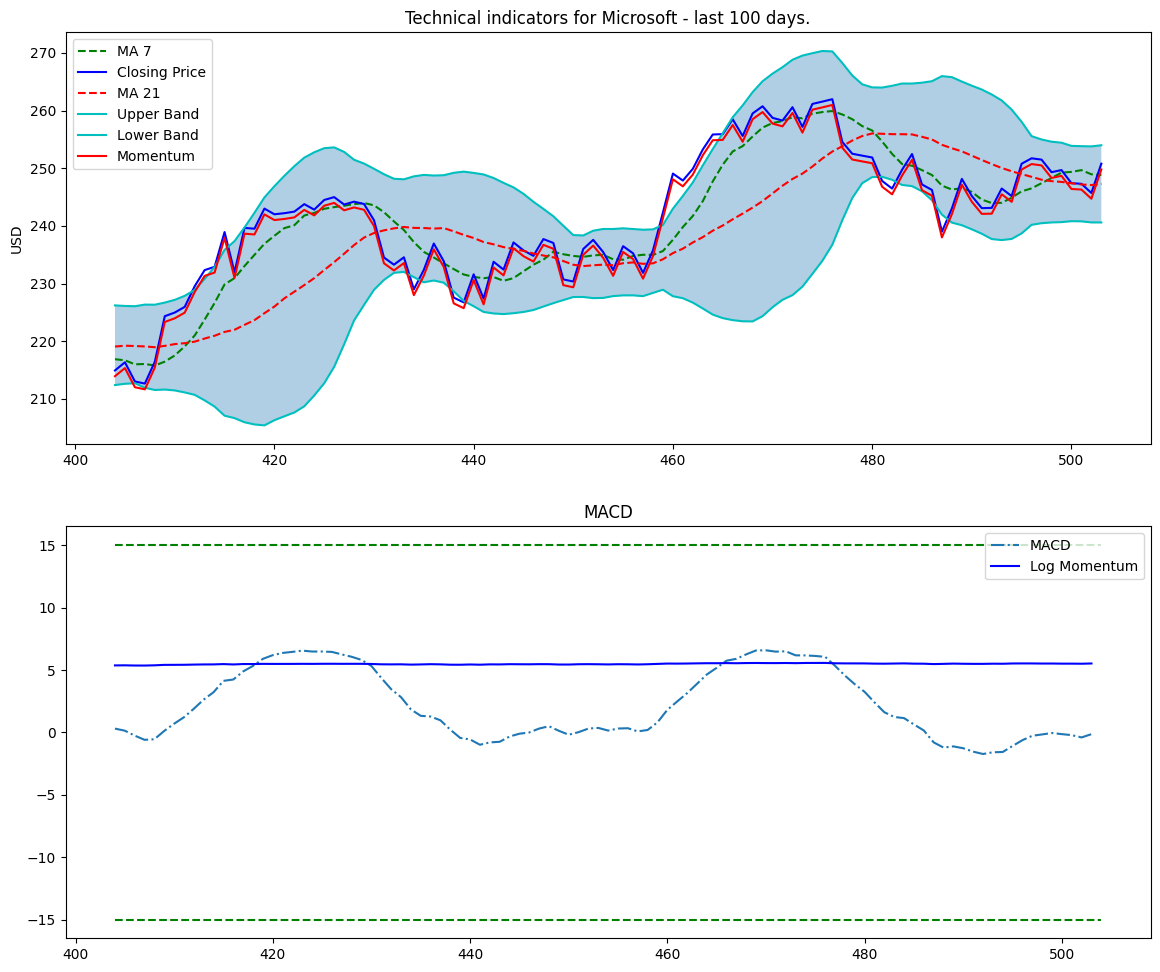}
 \caption{Plot of technical indicators on the Microsoft stock.}
 \end{figure}

The four technical indicators mentioned above are implemented to generate the labels ($\theta= 0, \pm1$) for the stock data sets.

\begin{itemize}
\item \textbf{MACD:} This indicator provides two outputs- MACD and MACD Signal Line (MACDSL). The instance when the MACD crosses the MACDSL for the first time, it is considered a sell ($-1$) signal, and for all consecutive instances when the MACD is above the MACDSL, it is considered to be a hold ($0$). Similarly, the moment it drops below the MACDSL it is considered to be a buy ($+1$) signal, and for all the following consecutive instances when it is below MACDSL, it is considered to be a hold ($0$). 
\item \textbf{RSI:}  RSI close value ranges from $0$ through $100$. A value greater than $70$ is considered to be a sell ($-1$) signal, value less than $30$ is considered to be a buy ($+1$) signal. If the value lies in between $30$ and $70$, it is a hold ($0$) signal.
\item \textbf{TRIX:}  This indicator provides a single TRIX output that oscillates about zero line. The instance when TRIX crosses the zero-signal line for the first time is considered to be a sell ($-1$) signal, and for all consecutive instances when TRIX is above zero signal line, it is considered to be a hold ($0$). Similarly, the moment it drops below zero signal line it is considered to be a buy ($+1$) signal, and for all the following consecutive instances when it is below zero signal line, it is considered to be a hold ($0$).
\item \textbf{Bollinger Bands:} We get two outputs from this indicator: upper band (UBB) and lower band (LBB). If the close value is greater than UBB, it is a sell ($-1$) signal, and if the value is less than LBB it is a buy ($+1$) signal. If the close value is in between the aforementioned two values, it is considered to be a hold ($0$).
\end{itemize}

The features used for time series forecasting are the values of close, momentum, and volatility. We use the latest $30$ days data to generate $7$ days of forecast. The indicator results and their measures are used as features in the label classification pipeline, i.e., the classification model is trained using indicator values as labels. The target for the final classification model is generated by comparing the present value of close price (denoted by, $\text{close}(t)$) with the value of $\text{close}(t-15)$, i.e., the close value $15$ days ago. For our analysis, we sell ($-1$) if the difference is positive and greater than $10\%$ of $\text{close}(t-15)$. We buy ($+1$) if the difference is negative and lesser than $10\%$ of $\text{close}(t-15)$. We hold ($0$) when the difference is between $0.1\text{close}(t-15)$ and $-0.1\text{close}(t-15)$. This parameter for holding could be tuned based on the required margin for the user. In summary,

\begin{equation}
\theta = +1, \quad \text{if} \quad \text{close}(t) - \text{close}(t-15) > 0.1 \text{close}(t-15),
\end{equation}
\begin{equation}
\theta = 0, \quad \text{if} \quad -0.1\text{close}(t-15) \leq (\text{close}(t) - \text{close}(t-15)) \leq 0.1\text{close}(t-15),
\end{equation}
\begin{equation}
\theta = -1, \quad \text{if} \quad \text{close}(t) - \text{close}(t-15) < -0.1\text{close}(t-15).
\end{equation}

\subsection{Time series analysis}
\label{sec42}

We implement the following analysis on the labels generated in the Subsection \ref{sec41}.

\begin{itemize}
\item \textbf{ARIMA}: ARIMA stands for AutoRegressive Integrated Moving Average which comes under a class of model that captures a suite of different standard temporal structures in time series data. We fit an ARIMA model on the stock data and observe that the closing price is almost accurately predicted but there is a lag when there is a trend reversal of the data. This is visualized in Figure 7 for the Microsoft data.

\begin{figure}[H]
\centering
\includegraphics[scale = 0.3]{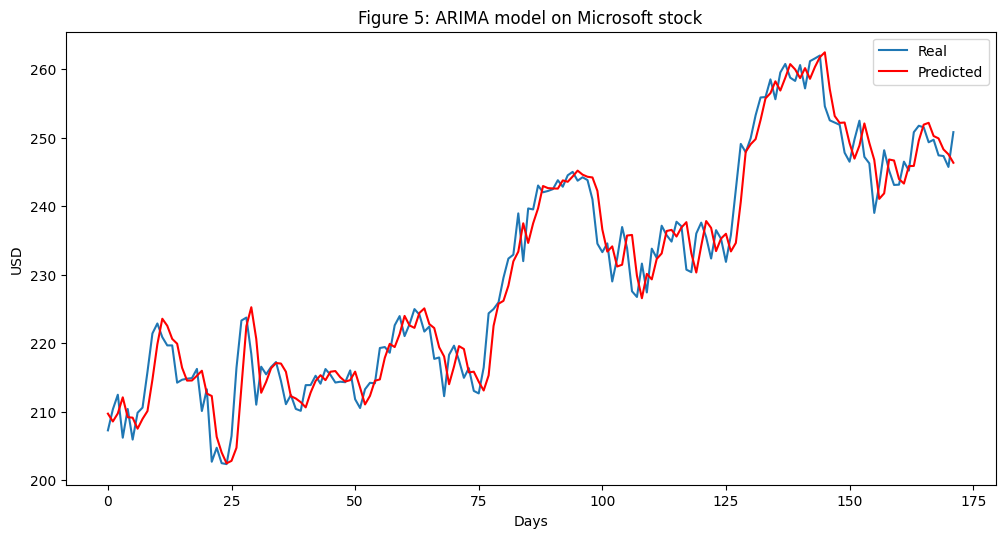}
\caption{ARIMA time series forecast on the Microsoft stock.}
\end{figure}   
        
\item \textbf{LSTM}: LSTM is one of the best time series prediction models. We use a single layer network with only one epoch initially to check whether it can catch any trend. It is observed that the values are not accurate, but it is able to capture the trend of the data. With this motivation, we decide to use a two-layer LSTM with $25$ epochs. The input is $30$ days of close data, and the output is the predicted close value. Finally, we concatenate and plot the close and predicted close values together. We find that the outputs are not identical, but the model is able to capture the trend.

\item \textbf{LSTM (multiple features)}:
        In this setting, we use a multi feature LSTM to include momentum, volatility, and close price of the stock data set. As more features are added, more information is captured by the model and better forecasting is observed. 

        The mathematical equations behind LSTM are as follows. If $X_t$ is the time series data, we define 
        $$g_t = \text{tanh}(X_t W_{xg} + h_{t-1} W_{hg} + b_g),$$
        
        $$i_t = \sigma(X_t W_{xi} + h_{t-1} W_{hi} + b_i),$$
        
        $$f_t = \sigma(X_t W_{xf} + h_{t-1} W_{hf} + b_f),$$
        
        $$o_t = \sigma(X_t W_{xo} + h_{t-1} W_{ho} + b_o),$$
        
        $$c_t = f_t \odot c_{t-1} + i_t \odot g_t,$$
        
        $$h_t = o_t \odot \text{tanh}(c_t),$$
        where $\odot$ is an element-wise multiplication operator, and, for all $x = [x_1, x_2, \ldots, x_k]^\top \in \mathbb{R}^k$ the two activation functions:
        $$\sigma(x) = \left[\frac{1}{1+\exp(-x_1)}, \ldots, \frac{1}{1+\exp(-x_k)}]\right]^\top,$$
        $$\text{tanh}(x) = \left[\frac{1-\exp(-2x_1)}{1+\exp(-2x_1)}, \ldots, \frac{1-\exp(-2x_k)}{1+\exp(-2x_k)}\right]^\top.$$
        
        The Time Series Data plots for LSTM with multiple features are shown below.
\end{itemize}

In Figure 8, a multi-feature LSTM model, based on close price, momentum, and volatility, is used to make a close price future forecast for the Microsoft stock. The blue and orange labels indicate training and validation sets, respectively. The green label indicates forecasts on validation set. In Figure 9, we observe the result on the Microsoft test set on a single day window for the year 2021.

\begin{figure}[H]
\centering
\includegraphics[scale = 0.6]{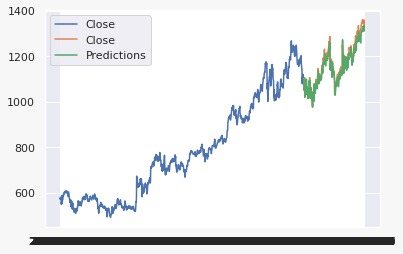}
\caption{Multi-feature LSTM on the Microsoft train data.}
\end{figure}

\begin{figure}[H]
\centering
\includegraphics[scale = 0.6]{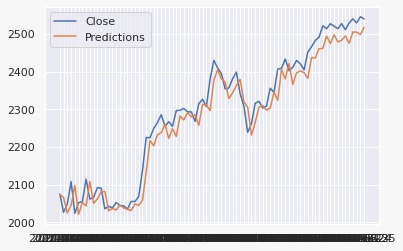}
\caption{Multi-feature LSTM on the Microsoft test data.}
\end{figure}

\subsection{Predictive pipeline}
\label{sec43}

On the generated outputs from Section \ref{sec42}, we perform label classification pipeline. 

\begin{itemize}
\item \textbf{Random forest}: Random forest is a supervised learning algorithm. It is an ensemble of decision trees, usually trained with the ``bagging” method, which is a combination of a lot of machine learning models to improve the final decision. Based on the close forecast generated for the next $7$ days in the window, we predict the signal on them using the supervised random forest classifier. The features in consideration are close, sign of moving average convergence divergence, Bollinger bands, triple exponential average and relative strength index. Also, the features associated with these indicators are also used. The classifier is trained on the data of past $30$ days.
\item \textbf{Boosted tree algorithms}: Boosting means combining a learning algorithm in series to achieve a strong learner from many sequentially connected weak learners. Trees in boosting are weak learners but adding many trees in series and each focusing on the errors from previous one make boosting a highly efficient and accurate model. There are many ways of iteratively adding learners to minimize a loss function. Common hyper-parameters for such models are maximum depth, maximum features, and minimum samples per leaf.
\item \textbf{SVM}: In machine learning, support-vector machine(SVM) is a supervised learning model with associated learning algorithms that analyze data for classification and regression analysis. SVM constructs a hyperplane or set of hyperplanes in a high-dimensional space, which can be used for classification, regression, or other tasks like outliers' detection. Intuitively, a good separation is achieved by the hyperplane that has the largest distance to the nearest training-data point of any class (so-called functional margin). In general, the larger the margin, the lower the generalization error of the classifier.
\end{itemize}

    The label classification model is based on a random forest classifier. There is a total of $13$ features for the classification. We already know how the target variable is defined. The $13$ features are \emph{close}, \emph{20-day simple moving average}, and \emph{MACD measures} (MACD sign, MACD signal line, MACD actual line and the difference between signal line and actual line), \emph{Bollinger Band measures} (BB sign, Upper BB and lower BB), \emph{RSI measures} (RSI sign, RSI close) and \emph{TRIX measures} (TRIX sign and TRIX line).

\subsection{Architecture}
\label{sec44}

In this subsection we combine the results in the Subsection \ref{sec41}, Subsection \ref{sec42}, and Subsection \ref{sec43}.  We implement a LSTM based architecture to forecast the time series data. For this, a classification model (for example, the random forest classifier) on top of the LSTM model is implemented to estimate the trend class. The forecasting model has $3$ LSTM layers and a dense layer which provides the estimated time series data. This data is then fed into the random forest classifier imported from \emph{sklearn} (in Python) with the default hyper-parameter setting for getting the class labels $+1$, $0$, and $-1$. The architecture is schematically shown in the Figure 10.

\begin{figure}[H]
\centering
 \includegraphics[scale = 0.45]{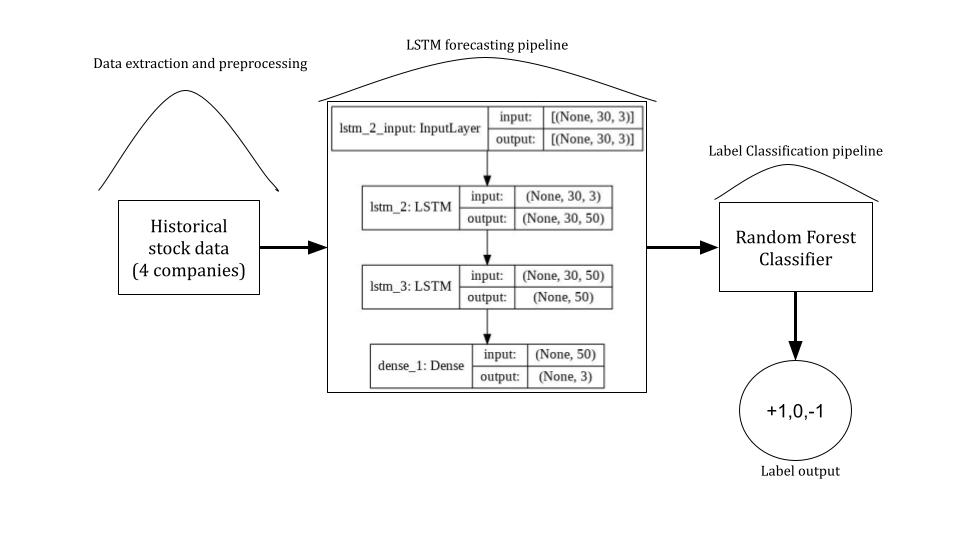}
\caption{Predictive pipeline architecture.}
\end{figure}

\section{Analysis of the results}
\label{sec5}

In this section, we propose certain tests in order to examine the goodness of fit of the forecasts generated by the proposed model and architecture. After that, we evaluate the classification task with some commonly used metrics.

\subsection{Statistical Analysis} 
 In order to check the goodness of fit, we used two tests- KS test, and KL divergence test. We observe that the data generated by the architecture in Subsection \ref{sec44} is a good estimated fit to the original data.
\begin{itemize}
\item \textbf{KS test}: The Kolmogorov–Smirnov test (KS test) is a non-parametric test of the equality of continuous, one-dimensional probability distributions that can be used to compare a sample with a reference probability distribution (one-sample KS test), or to compare two samples (two-sample KS test). We have used the test on our time series forecast to see if the predictions and actual values of close come from the same distribution. 
\item \textbf{KL divergence test}: To measure the difference between two probability distributions over the same variable $X$, a measure, called the Kullback-Leibler (KL) divergence test, has been popularly used in the data mining literature. Here we assume that the original distribution of time series data is $P(x)$ and the predicted distribution is $Q(x)$.
 For discrete probability distributions $P$ and $Q$ defined on the same probability space $X$, the relative entropy from $Q$ to $P$ is defined as:
        $$D_{KL}(P||Q) = \sum_{x \in X} P(x) \log \frac{P(x)}{Q(x)}.$$
\end{itemize}

The results of goodness of fit are shown in the Table 1 for all the four data sets.

\begin{table}[H]
\centering
\caption{Goodness of fit.}
  \begin{tabular}{ | l | c | r |}
    \hline
           Stock & KS-Test ($p$-value) & KL Div Test (entropy) \\  \hline \hline
           \textbf{Apple} & 0.47 & 3.99e-05 \\  \hline
            \textbf{Amazon} & 0.87 & 0.0001\\ \hline
          \textbf{Google} & 0.15 & 3.01e-05 \\ \hline
        \textbf{Microsoft} & 0.99 & 6.54e-05 \\ \hline
  \end{tabular}
\end{table}

\subsection{Classification results}

\begin{itemize}
\item \textbf{Weighted average F1-score}:  F1-score is a classification metric. It is calculated from the precision and recall score for a case. Precision provides the relevance of the selection and recall (also known as sensitivity) provides the number of relevant items considered. For the weighted average F1-score, we calculate the F1-score for each label and obtain the average considering the proportion for each label in the data set.
\item \textbf{Accuracy}: Accuracy is also a common metric for evaluating classification models. It is defined as the ratio of correct predictions over total prediction. An accurate model in a classification model must be tested for its sensitivity for better generalization.
\end{itemize}
Based on the tree-based predictive model, we obtain the following result: $7$-days forecast is generated from the LSTM forecasting network using the previous $30$-days of data. This method is implemented to obtain predicted close prices for $4$ windows, each of $7$ days for June 2021. This predicted close value is fed into the label classification pipeline. For all the stocks in our consideration we obtain the accuracy and F1-scores. The results, shown in Table 2, seem to be quite promising and consistent with the current benchmarks. These scores are the performance value of the predictive model for coming up with an accurate call to hold, sell or buy the stock. F1-score is measured to understand the general specificity of the model. A wrong prediction may incur a lot of loss to the trader. Consequently, a good performance on both the aspects is very important.

\begin{table}[H]
\centering
\caption{Classification results.}
  \begin{tabular}{ | l | c | r |}
    \hline
Stock & Accuracy & F1-score \\  \hline \hline
 \textbf{Apple} & 91.66 & 0.91 \\  \hline
        \textbf{Amazon} & 95.8 & 0.95 \\ \hline
        \textbf{Google} & 95.83 & 0.92 \\ \hline
     \textbf{Microsoft} & 95.80 & 0.94  \\ \hline
  \end{tabular}
\end{table}

Table 3 shows the accuracies and F1-scores on each of the features used (i.e., Bollinger Bands, MACD, TRIX and RSI) for the four data sets (Microsoft (MSFT), Google (goog), Apple (apple), and Amazon (amazon)).

\begin{table}[H]
\centering
\caption{Accuracies and F1-scores on each of the features.}
    \includegraphics[scale = 0.7]{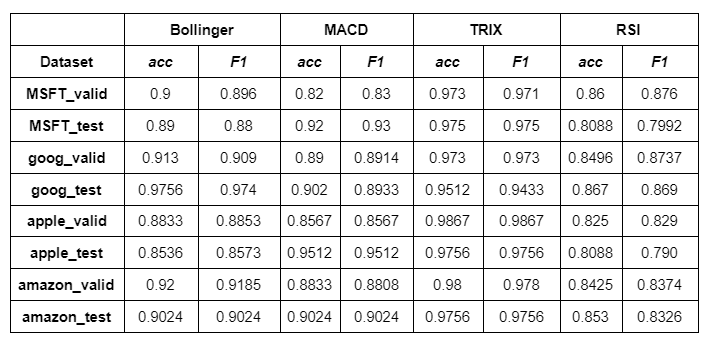}
\end{table}

Tables 4, 5, 6, and 7 show the accuracies and F1-scores for different types for machine learning classification model. To validate the final results more strongly we consider two different test sets for testing out all the label classification algorithms. Tables 4 and 5 show the results for the first test set from 02/22/21 (February 22, 2021) to 04/26/21 (April 26, 2021). Tables 6 and 7 show the results for the second test set from 04/27/21 (April 27, 2021) to 06/25/21 (June 25, 2021).

\begin{table}[H]
\centering
\caption{Accuracy results on test set 1: 02/22/21 to 04/26/21.}
  \begin{tabular}{ | l | c | c | c |}
    \hline
  Data set & Random Forest & SVM Classifier & XGB Classifier \\  \hline \hline
 \textbf{Google test} & 88 & 90 & 79\\  \hline
        \textbf{Microsoft test} & 88 & 90 & 81 \\ \hline
        \textbf{Amazon test} & 90 & 86 & 88\\ \hline
     \textbf{Apple test} & 85 & 86  & 82\\ \hline
  \end{tabular}
\end{table}

\begin{table}[H]
\centering
\caption{F1-score results on test set 1: 02/22/21 to 04/26/21.}
\begin{tabular}{ | l | c | c | c |}
\hline
Data set & Random Forest & SVM Classifier & XGB Classifier \\  \hline \hline
\textbf{Google test} & 0.83 & 0.91 & 0.75\\  \hline
    \textbf{Microsoft test} & 0.83 & 0.91 & 0.83 \\ \hline
    \textbf{Amazon test} & 0.86 & 0.84 & 0.85\\ \hline
 \textbf{Apple test} & 0.84 & 0.85  & 0.79\\ \hline
\end{tabular}  
\end{table}

\begin{table}[H]
\centering
\caption{Accuracy results on test set 2: 04/27/21 to 06/25/21.}
  \begin{tabular}{ | l | c | c | c |}
    \hline
  Data set & Random Forest & SVM Classifier & XGB Classifier \\  \hline \hline
 \textbf{Google test} & 98 & 97 & 97\\  \hline
        \textbf{Microsoft test} & 98 & 98 & 96 \\ \hline
        \textbf{Amazon test} & 97 & 98 & 97\\ \hline
     \textbf{Apple test} & 93 & 98  & 93\\ \hline
  \end{tabular}
\end{table}

\begin{table}[H]
\centering
\caption{F1-score results on test set 2: 04/27/21 to 06/25/21.}
\begin{tabular}{ | l | c | c | c |}
\hline
 Data set & Random Forest & SVM Classifier & XGB Classifier \\  \hline \hline
\textbf{Google test} & 0.97 & 0.97 &0.97\\  \hline
    \textbf{Microsoft test} & 0.98 & 0.98 & 0.97 \\ \hline
    \textbf{Amazon test} & 0.97 & 0.97 & 0.97\\ \hline
 \textbf{Apple test} & 0.96 & 0.98  & 0.97\\ \hline
\end{tabular}  
\end{table}

The analysis above corroborates the idea that the architecture proposed in the paper can accurately capture the short-term movement of one of the four concerned stocks. This is important for short-term traders. In addition, this also improves the underlying stochastic model by finding the value of $\theta$ ($0$, or $\pm1$) that can be implemented to \eqref{Btheta} and \eqref{rhotheta}. The resulting model can be considered as an improved BN-S model.

\section{Conclusion}
\label{sec6}

In this work, we present a model that captures the trend for four different stock data sets, namely, Amazon, Apple, Google, and Microsoft. A thorough exploratory data analysis gives us an intuition of the data, and the hidden clusters provide us the idea about the three classes- $-1$, $0$, and $+1$. Furthermore, we incorporate momentum and volatility in stock market analysis with deep learning-based forecasting, followed by the use of technical indicators based on momentum, volatility, and trend for the trend classification task. The model proves to be robust in predicting the direction of the stock movement. This is also validated from good accuracy and F1-score. For all the data sets, our final accuracy lies in the range of 90-95\% on a 7-day prediction window. The $p$-values obtained from the KS test and the entropy of KL divergence test also show the robustness of LSTM network employed for forecasting the time series stock data.  

For the future work, we consider using ensemble modeling for the classification pipeline and experimenting with different neural network architectures. Moreover, in our future work we plan to incorporate features like those extracted from sentiment analysis of the companies. In addition, it may be possible that the analysis presented in the paper is implementable to other stocks.

\appendix

\section{Appendix: statistical quantities for the data sets}
\label{apendixA}

\begin{itemize}
\item \textbf{Amazon Data}
\end{itemize}
    
    \begin{center}
        \scalebox{0.75}{
        \begin{tabular}{|c c c c c c c c c c c c|} 
             
            \hline
             & \textbf{Open} & \textbf{High} & \textbf{Low} & \textbf{Close} & \textbf{Adj Close} & \textbf{Volume} & \textbf{SMA close} & \textbf{EMA close} & \textbf{up} & \textbf{down} & \textbf{RSI close} \\ 
            \hline \hline
            \textbf{mean} & 984.3 & 993.2 & 973.6 & 983.9 & 983.9 & 4.17e+06 & 983.5 & 979.2 & 6.51 & -5.55 & 56.15 \\ 
            \hline
            \textbf{std} & 566.8 & 571.6 & 560.6 & 566.3 & 566.3 & 2.29e+06 & 565.5 & 564.4 & 12.81 & 12.99 & 16.81 \\
            \hline
            \textbf{min} & 284.4 & 290.4 & 284.0 & 286.9 & 286.9 & 8.81e+05 & 291.1 & 294.5 & 0.00 & -139.35 & 11.43 \\
            \hline
            \textbf{25\%} & 439.3 & 444.7 & 435.5 & 439.4 & 439.4 & 2.72e+06 & 439.5 & 435.8 & 0.00 & -5.22 & 44.36 \\
            \hline
            \textbf{50\%} & 818.0 & 821.6 & 812.5 & 817.8 & 817.8 & 3.56e+06 & 818.8 & 816.1 & 0.90 & 0.00 & 55.94\\ 
            \hline
            \textbf{75\%} & 1604.0 & 1622.7 & 1590.7 & 1602.9 & 1602.9 & 4.80e+08 & 1603.2 & 1600.4 & 8.16 & 0.00 & 68.25\\
            \hline
            \textbf{max} & 2038.1 & 2050.5 & 2013.0 & 2039.5 & 2039.5 & 2.38e+07 & 2012.1 & 1985.2 & 128.51 & 0.00 & 93.97\\
            \hline
            
        \end{tabular}}
    
    \end{center}
    
    \begin{itemize}
        \item \textbf{Apple Data}
    \end{itemize}

    \begin{center}
        \scalebox{0.75}{
        \begin{tabular}{|c c c c c c c c c c c c|} 
             
            \hline
             & \textbf{Open} & \textbf{High} & \textbf{Low} & \textbf{Close} & \textbf{Adj Close} & \textbf{Volume} & \textbf{SMA close} & \textbf{EMA close} & \textbf{up} & \textbf{down} & \textbf{RSI close} \\ 
            \hline \hline
            \textbf{mean} & 35.98 & 36.29 & 35.67 & 35.99 & 34.20 & 1.61e+08 & 35.97 & 35.82 & 0.21 & -0.17 & 56.46 \\ 
            \hline
            \textbf{std} & 11.64 & 11.74 & 11.55 & 11.66 & 12.03 & 9.06e+07 & 11.57 & 11.48 & 0.35 & 0.36 & 18.08 \\
            \hline
            \textbf{min} & 17.68 & 17.91 & 17.62 & 17.84 & 15.87 & 4.54e+07 & 17.92 & 18.45 & 0.00 & -3.93 & 8.21 \\
            \hline
            \textbf{25\%} & 26.96 & 27.21 & 26.70 & 26.98 & 24.89 & 1.01e+08 & 27.05 & 27.02 & 0.00 & -0.20 & 43.06 \\
            \hline
            \textbf{50\%} & 32.29 & 32.62 & 32.07 & 32.34 & 29.60 & 1.36e+08 & 32.30 & 32.05 & 0.02 & 0.00 & 56.77\\ 
            \hline
            \textbf{75\%} & 43.81 & 44.29 & 43.62 & 43.95 & 42.51 & 1.97e+08 & 43.92 & 43.64 & 0.31 & 0.00 & 69.88\\
            \hline
            \textbf{max} & 72.77 & 73.49 & 72.02 & 72.87 & 72.02 & 1.06e+09 & 71.97 & 70.82 & 2.80 & 0.00 & 95.93\\
            \hline
            
        \end{tabular}}
    
    \end{center}
    
    \begin{itemize}
        
        \item \textbf{Google Data}
    
    \end{itemize}
    
    \begin{center}
        \scalebox{0.75}{
        \begin{tabular}{|c c c c c c c c c c c c|} 
             
            \hline
             & \textbf{Open} & \textbf{High} & \textbf{Low} & \textbf{Close} & \textbf{Adj Close} & \textbf{Volume} & \textbf{SMA close} & \textbf{EMA close} & \textbf{up} & \textbf{down} & \textbf{RSI close} \\ 
            \hline \hline
            \textbf{mean} & 854.1 & 861.3 & 846.7 & 854.2 & 854.2 & 1.83e+06 & 853.9 & 851.6 & 4.59 & -4.07 & 54.43 \\ 
            \hline
            \textbf{std} & 248.3 & 250.6 & 246.5 & 248.7 & 248.7 & 1.06e+06 & 247.9 & 246.9 & 8.22 & 8.18 & 15.90 \\
            \hline
            \textbf{min} & 493.3 & 494.6 & 486.2 & 491.2 & 491.2 & 7.92e+03 & 496.1 & 503.4 & 0.00 & -99.09 & 15.37 \\
            \hline
            \textbf{25\%} & 601.6 & 604.1 & 595.4 & 599.6 & 599.6 & 1.23e+06 & 604.4 & 598.3 & 0.00 & -5.13 & 42.90 \\
            \hline
            \textbf{50\%} & 798.2 & 803.5 & 793.3 & 797.1 & 797.1 & 1.53e+06 & 798.8 & 792.9 & 0.42 & 0.00 & 53.79\\ 
            \hline
            \textbf{75\%} & 1083.5 & 1094.2 & 1073.4 & 1082.8 & 1082.8 & 2.05e+06 & 1083.4 & 1083.1 & 6.75 & 0.00 & 65.35\\
            \hline
            \textbf{max} & 1363.3 & 1365.0 & 1352.7 & 1361.2 & 1361.2 & 1.11e+07 & 1354.9 & 1348.9 & 118.3 & 0.00 & 99.16\\
            \hline
            
        \end{tabular}}
    
    \end{center}
    
    \begin{itemize}
        \item \textbf{Microsoft Data}
    \end{itemize}
    
    \begin{center}
        \scalebox{0.75}{
        \begin{tabular}{|c c c c c c c c c c c c|} 
             
            \hline
             & \textbf{Open} & \textbf{High} & \textbf{Low} & \textbf{Close} & \textbf{Adj Close} & \textbf{Volume} & \textbf{SMA close} & \textbf{EMA close} & \textbf{up} & \textbf{down} & \textbf{RSI close} \\ 
            \hline \hline
            \textbf{mean} & 74.55 & 75.14 & 73.90 & 74.56 & 70.39 & 2.98e+07 & 74.50 & 74.17 & 0.40 & -0.32 & 56.83 \\ 
            \hline
            \textbf{std} & 32.60 & 32.81 & 32.30 & 32.58 & 33.34 & 1.44e+07 & 32.43 & 32.27 & 0.72 & 0.69 & 15.06 \\
            \hline
            \textbf{min} & 34.73 & 35.88 & 34.63 & 34.98 & 30.11 & 7.42e+06 & 35.61 & 36.11 & 0.00 & -6.09 & 15.71 \\
            \hline
            \textbf{25\%} & 46.93 & 47.45 & 46.54 & 47.00 & 41.99 & 2.12e+07 & 46.94 & 46.73 & 0.00 & -0.36 & 46.29 \\
            \hline
            \textbf{50\%} & 62.70 & 63.08 & 62.27 & 62.63 & 58.45 & 2.66e+07 & 62.71 & 62.64 & 0.05 & 0.00 & 56.26\\ 
            \hline
            \textbf{75\%} & 101.1 & 101.8 & 99.51 & 101.1 & 97.39 & 3.40e+07 & 100.8 & 100.7 & 0.57 & 0.00 & 67.25\\
            \hline
            \textbf{max} & 159.4 & 159.6 & 158.2 & 158.9 & 156.6 & 2.02e+08 & 158.0 & 156.5 & 6.59 & 0.00 & 99.11\\
            \hline
            
        \end{tabular}}
    
    \end{center}

\end{document}